\documentclass[5p, twocolumn]{elsarticle}

\usepackage{amssymb}
\usepackage{lineno}
\usepackage{multirow}
\usepackage{graphicx}
\usepackage{subfigure}
\usepackage{amssymb, amsmath, bm}
\usepackage{mathtools}
\usepackage{tensor}
\usepackage{mathrsfs}
\usepackage{booktabs}
\usepackage{url}
\usepackage{color}
\journal{Computer Methods and Programs in Biomedicine}

\begin{document}

\begin{frontmatter}

\title{Hybrid Attention for Automatic Segmentation of Whole Fetal Head in \\Prenatal Ultrasound Volumes}

\author[Affil1]{Xin Yang}
\author[Affil2,Affil3]{Xu Wang}
\author[Affil2,Affil3]{Yi Wang}
\author[Affil2,Affil3]{Haoran Dou}
\author[Affil4]{Shengli Li}
\author[Affil4]{Huaxuan Wen}
\author[Affil4]{Yi Lin}
\author[Affil1]{Pheng-Ann~Heng}
\author[Affil2,Affil3]{Dong~Ni\corref{cor1}}

\address[Affil1]{Department of Computer Science and Engineering, The Chinese University of Hong Kong, Hong Kong, China.}
\address[Affil2]{National-Regional Key Technology Engineering Laboratory for Medical Ultrasound, School of Biomedical Engineering, Health Science Center, Shenzhen University, Shenzhen, China.}
\address[Affil3]{Medical UltraSound Image Computing (MUSIC) Lab, Shenzhen University, Shenzhen, China.}
\address[Affil4]{Department of Ultrasound, Affiliated Shenzhen Maternal and Child Healthcare Hospital of Nanfang Medical University, Shenzhen, China.}
% Replace capitalized text with the appropriate information (use standard capitalization rules for your text, not all capitals.
\cortext[cor1]{Corresponding Author: Dong Ni; Rm A2-523, Health Science Center, Shenzhen University, Shenzhen, Guangdong Province, China, 518037; \textit{nidong@szu.edu.cn}; (+86)755-86671920.}

\begin{abstract}
	\textit{Background and Objective:} Biometric measurements of fetal head are important indicators for maternal and fetal health monitoring during pregnancy. 3D ultrasound (US) has unique advantages over 2D scan in covering the whole fetal head and may promote the diagnoses. However, automatically segmenting the whole fetal head in US volumes still pends as an emerging and unsolved problem. The challenges that automated solutions need to tackle include the poor image quality, boundary ambiguity, long-span occlusion, and the appearance variability across different fetal poses and gestational ages. In this paper, we propose the first fully-automated solution to segment the whole fetal head in US volumes. \par
	
	\textit{Methods:} The segmentation task is firstly formulated as an end-to-end volumetric mapping under an encoder-decoder deep architecture. We then combine the segmentor with a proposed hybrid attention scheme (HAS) to select discriminative features and suppress the non-informative volumetric features in a composite and hierarchical way. With little computation overhead, HAS proves to be effective in addressing boundary ambiguity and deficiency. To enhance the spatial consistency in segmentation, we further organize multiple segmentors in a cascaded fashion to refine the results by revisiting context in the prediction of predecessors. \par
	
	\textit{Results:} Validated on a large dataset collected from 100 healthy volunteers, our method presents superior segmentation performance (DSC (Dice Similarity Coefficient), 96.05\%), remarkable agreements with experts (-1.6$\pm$19.5 mL). With another 156 volumes collected from 52 volunteers, we ahieve high reproducibilities (mean standard deviation 11.524 mL) against scan variations. \par
	
	\textit{Conclusion:} This is the first investigation about whole fetal head segmentation in 3D US. Our method is promising to be a feasible solution in assisting the volumetric US-based prenatal studies.
\end{abstract}


\end{frontmatter}

% \linenumbers

\section{Introduction}
\label{intro}
Prenatal examinations during different trimesters depend heavily on ultrasound (US) scanning, which is well-recognized as real-time, non-invasive and radiation-free. Biometric measurements interpreted from US images are foundations for the evaluation of fetal and maternal health across different gestational ages \citep{hadlock1985estimation}. \par

\begin{figure}[h]
	\centering
	\includegraphics[width=1.0\linewidth]{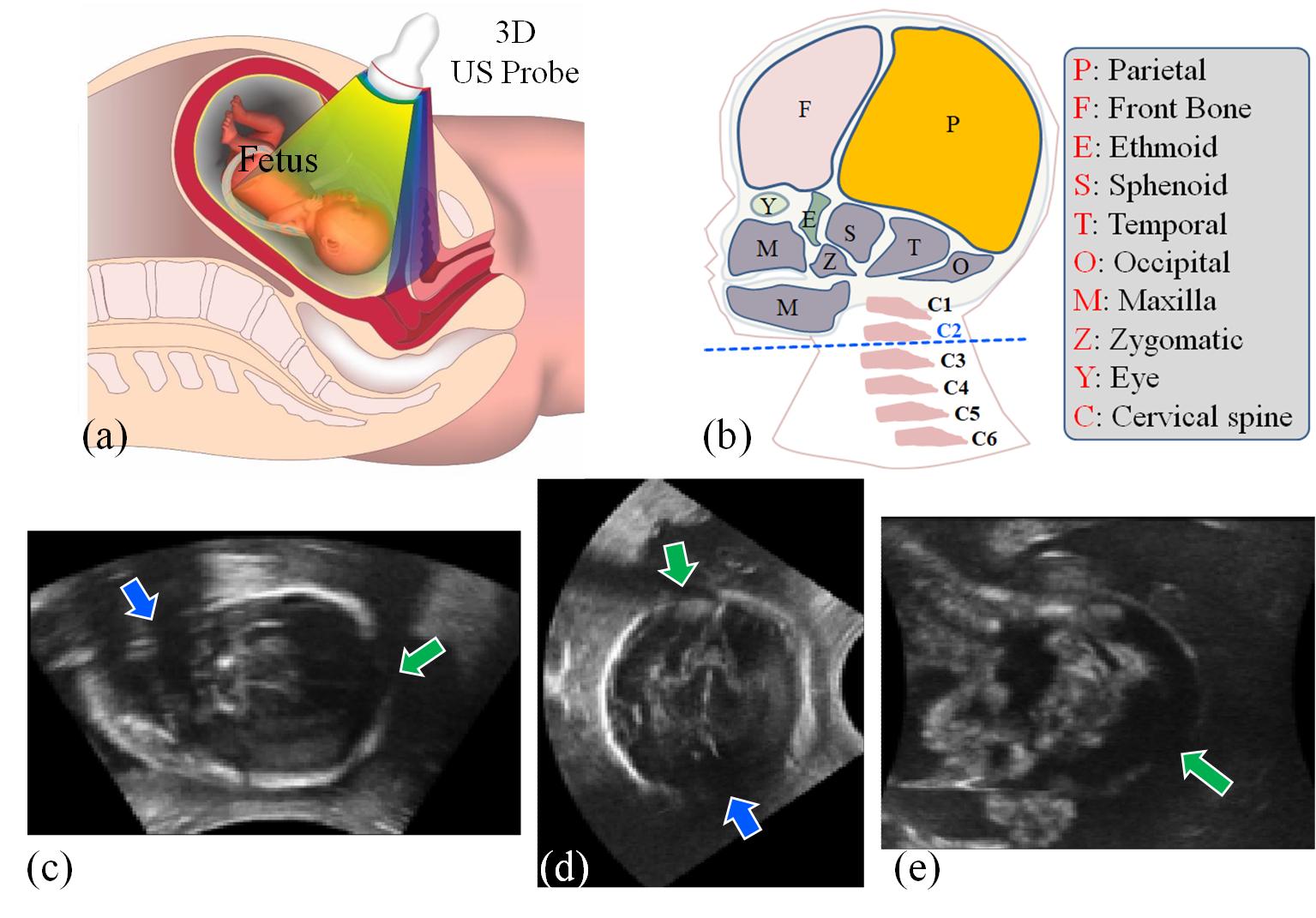}
	\caption{Volumetric ultrasound of the whole fetal head. (a) Illustration of the 3D ultrasound imaging process around the fetal head. Varying fetal poses are allowed during our acquisition. (b) Anatomical definition of the whole fetal head region (area above the blue line). Skull (includes \textit{P}, \textit{F}, \textit{E}, \textit{S}, \textit{T} and \textit{O}) is only a sub-region of the whole head. (c)-(e) Coronal, traverse and sagittal planes from a fetal head ultrasound volume. Arrows denote the various occluded sites (blue arrows), deficient and ambiguous boundaries (green arrows) around the fetal head.}
	\label{fig:challenges}
\end{figure}

Among all the biometrics, measurements focusing on fetal head are major indicators accepted by sonographers, which are explicit in reflecting the growth stage of fetus. By combining the measurements of fetal head with that of other anatomical structures, like fetal abdomen and femur, sonographers can further estimate the fetal weight and hence gain better insights for diagnosis. However, limited by 2D US scanning, current clinical measurement pipeline often exposes the diagnosis of fetal head to potential risks. First, 2D measurements obtained from approximated geometry primitives, like line and ellipse, are obviously rough in describing the complex 3D geometry of fetal head. Sonographers often need multiple 2D measurements, like Head Circumference (HC) and Biparietal Diameter (BPD), to justify their diagnoses \cite{rueda2014evaluation}. Second, selecting standard planes which contain key anatomical substructures is a prerequisite for measurement. Bias of sonographers in this selection step often enlarges the discrepancy in diagnoses \cite{Ni_umb_plane}. Previous automatic solutions \cite{yu2008fetal,wu2017cascaded,li2018automatic} for fetal head measurement partially address the problems, but they are still limited by the 2D US imaging. \par

% advantages of 3D US
As shown in Fig. \ref{fig:challenges}, 3D US is emerging as a promising alternative in circumventing the aforementioned problems \cite{Tutschek}. It provides a broad volumetric field of view and therefore enables sonographers to inspect the fetal head anatomy in an ever straightforward way (Fig. \ref{fig:challenges}(a)). Volumetric scanning is less expert-dependent than 2D scanning and thus alleviates the risk of discrepancy in image acquisition. Biometric measurements extracted from 3D US, such as volume, are more representative and comprehensive than 2D ones for diagnosis. Volumetric metrics may also provide earlier indicators than planar ones for prognosis \cite{Tutschek}. \par

Although 3D US is attractive for fetal head imaging, efficient and effective tools for whole fetal head segmentation and quantitative analysis running on the massive volume are still absent. As illustrated in Fig. \ref{fig:challenges}(c)-(e), automated segmentation solutions need to tackle the following challenges possessed by volumetric US of whole fetal head: \textit{i)} the poor image quality resulting from the speckle noise and low resolution, \textit{ii)} inevitable boundary ambiguity caused by low tissue contrast and long-span shadow occlusion caused by the severe acoustic attenuation on skull, particularly in the far field, \textit{iii)} the large varieties of fetal head in appearance (particularly the inner structures), scale, orientation and shape across different fetal poses and gestational ages. \par

% related work
Much work have been proposed for prenatal volumetric US segmentation. Semi-automatic solutions, like VOCAL (GE Healthcare), were investigated in clinic to segment fetal anatomical structures \cite{Luewan_vocal_placentaV}. These semi-automatic solutions often simplify the segmentation and discard many important details. In \cite{Dahdouh}, Dahdouh \textit{et al.} explored both intensity distributions and shape priors to segment fetus. Feng \textit{et al.} exploited boundary traces to extract fetal limb volume in \cite{Feng_limb}. Recently, Namburete \textit{et al.} proposed a 3D deformable parametric surface to represent and fit fetal skull for fetal brain maturity evaluation \cite{Namburete_head}. Although shape models provide proper constraints for robust fitting, they are initialization-dependent and rough in capturing the case-specific boundary details. Traditional machine learning methods, like random forests, were leveraged to segment fetal brain structures \cite{Yaqub_brain}. Structured Random Forest was further used to segment fetal skull \cite{cerrolaza2017fetal}. \par

Deep neural networks (CNN) have drastically taken over the traditional methods in US image segmentation \cite{Shen_review,torrents2019segmentation}. Characterized with the end-to-end dense mapping, fully convolutional network (FCN) \cite{Long_fcn} was adopted by \cite{wu2017cascaded} for 2D prenatal US image segmentation with a high performance. A 3D FCN with Recurrent Neural Network was further developed by \cite{yang2017towards} to segment the whole fetus, placenta and gestational sac in early gestational ages. Namburete \textit{et al.} \cite{Namburete2018fully} used the shape model to generate skull masks and trained deep neural networks to segment fetal skull in 3D US volumes. They reported the average segmentation DSC as 83$\%$. Recently, Cerrolaza \textit{et al.} \cite{cerrolaza2018deep} proposed to combine the acoustic shadow casting map with a deep network to segment 3D fetal skull, also achieving an average DSC of 83$\%$. A deep conditional generative network was further developed for the 3D fetal skull reconstruction from few 2D slices by \cite{cerrolaza20183d}, reporting the average DSC of 90$\%$. \par

Although the tasks in \cite{cerrolaza2017fetal,Namburete2018fully,cerrolaza2018deep,cerrolaza20183d} are similar to the work in this paper, their focuses are the fetal skull segmentation, while our task is segmenting the whole fetal head. As shown in Fig. \ref{fig:challenges}(b), whole fetal head is defined as the region of fetus on top of the plane determined by the fetal lower jaw and cervical spine \textit{C2} (denoted as the blue line in Fig. \ref{fig:challenges}(b)). It not only includes the skull (\textit{P, F, E, S, T, O}), but also the maxillo-facial structures of fetus (\textit{M, Z, Y, C}). Segmentation of whole fetal head is more informative than skull segmentation for fetal growth evaluation. However, since the maxillo-facial areas of fetus often present larger non-rigid deformations, occlusions and boundary deficiencies than the skull (Fig. \ref{fig:challenges}(c)-(e)), our task is much more difficult than the skull segmentation. \par

\begin{figure*}[h]
	\centering
	\includegraphics[width=0.9\linewidth]{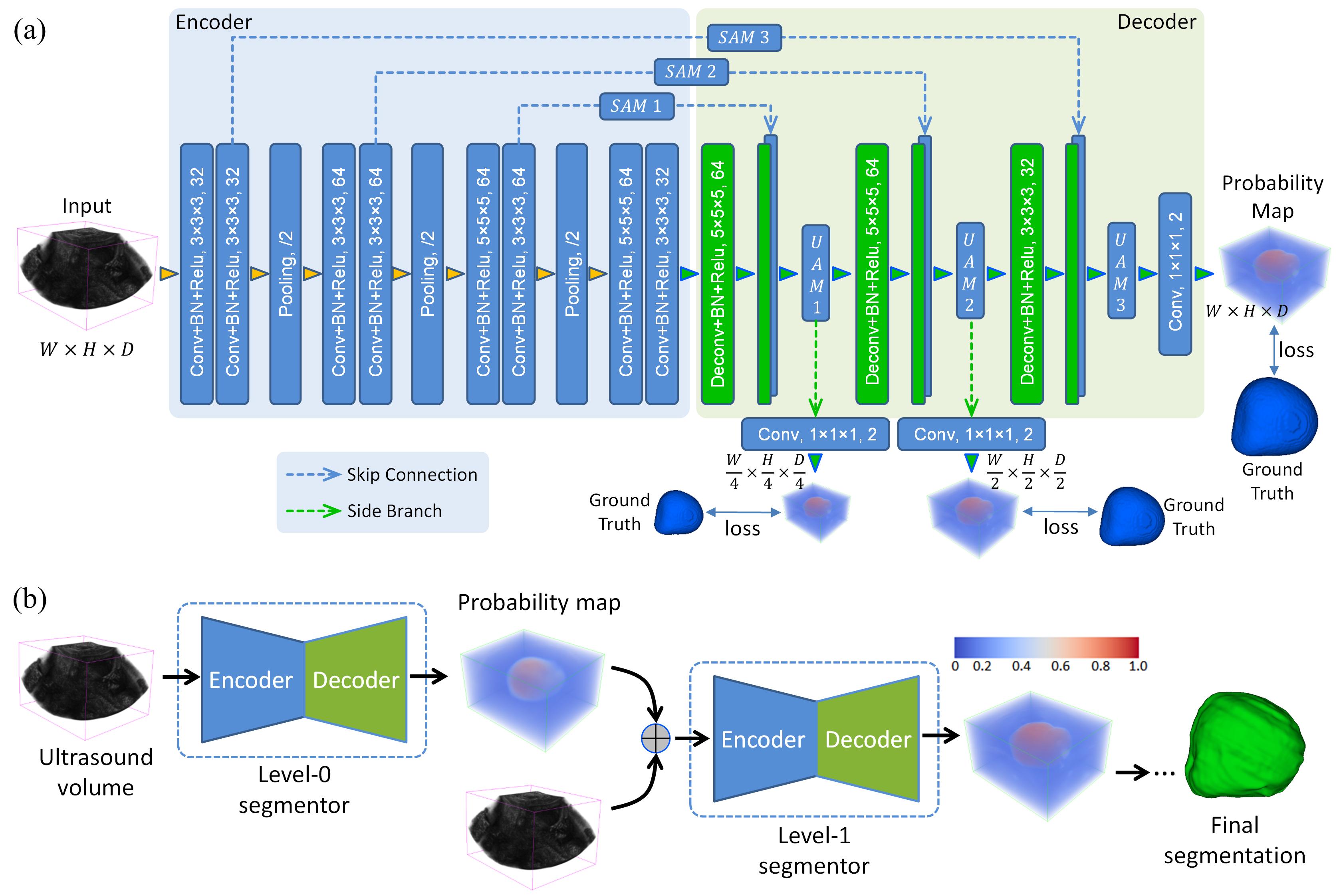}
	\caption{Schematic view of our segmentation solution. (a) Architecture of our proposed segmentor. Encoder-decoder design with 3D operators digests the whole US volume. Hybrid attention scheme is combined to enhance the feature maps in key sites. Deep supervision with side branch boosts the network training. (b) Cascaded segmentors in an Auto-Context framework for segmentation refinement. Summation of US volume and probability map of whole fetal head is the input of each context level.}
	\label{fig:framework}
\end{figure*}

% brief introduction of our paper
In this paper, we propose the first fully-automated solution to segment the whole fetal head in US volumes. In order to fully explore the context in the whole US volume, we firstly formulate the segmentation task as an end-to-end volumetric mapping under an encoder-decoder deep architecture. Making every feature descriptor to be representative is crucial for deep networks, especially for our volumetric network which suffers severe limitations from computation resources. Therefore, we propose a hybrid attention scheme (HAS) and combine it with our segmentor to promote discriminative features and suppress the non-informative volumetric features in a composite and hierarchical way. HAS brings minimal computation overhead and proves to be effective in helping segmentor combating boundary ambiguity and deficiency. To enhance the spatial consistency in the segmentation, we further organize multiple segmentors in a cascaded fashion to refine the results by explicitly revisiting the global shape context encoded in the predictions from predecessors. With experiments on a large dataset, our method presents the ability to tackle a wide range of gestational ages with superior segmentation performance (average DSC as 96$\%$), high agreements with experts and decent reproducibilities (mean standard deviation 11.524 mL). The automated segmentation will not only benefit the extraction of representative biometrics in fetal head, but also have potentials in boosting many advanced prenatal studies, like brain alignment \cite{Namburete2018fully}, volume stitching \cite{Gomez_3Dus_regis} and longitudinal analysis \cite{Namburete_head}. Code is publicly available \textcolor{blue}{\url{https://github.com/wxde/USegNet-DS-HAS}}. \par

%%%%%%%%%%% MATERIALS AND METHODS
\section{Materials and Methods}
\label{MaM}
\subsection{Datasets}
% Datasets
To cover the most important gestational ages where fetal head is intensively examined, we built a dataset consisting of 100 US volumes of fetal head acquired from 100 healthy pregnant women volunteers, with a gestational age (GA) ranging from 20 to 31 weeks. All the data acquisition have been approved by the local Institutional Review Board. All the volunteers that participated in this study have reviewed and signed the consent forms. All the volumes were anonymized and acquired by an experienced sonographer using an US machine (S50, SonoScape Medical Corp., Shenzhen, Guangdong Province, (China)) with an integrated 3D probe. The probe has a $75^{\circ}$ scan angle to ensure a complete scanning of the whole fetal head. Fetus is in static state during scanning and free fetal poses are considered during acquisition. Varying scanning orientation of US probe is accepted to ensure the acquisition quality. The original size of volume is 388$\times$258$\times$440 with a spacing of 0.38$\times$0.38$\times$0.38 $mm^{3}$. An expert with 10-year experience manually delineated all volumes as ground truth. Being skilled in using the annotation software \textit{ITK-SNAP} \cite{itk-snap}, the expert needs about 2 hours to finish the annotation of one volume. All the annotation results are double-checked by a senior expert with 20-year experience. We then randomly split the dataset into 50, 50 volumes for training and testing. Regarding the varying fetal head pose, the training dataset is firstly augmented to 600 with flipping and $180^{\circ}$-step rotation around three axes. We then augment the training data by applying the Random Erasing \cite{zhong2017random} to randomly erase a sub-region with 0 at a random position around the whole fetal head to mimic the ubiquitous acoustic shadow. Finally, the training dataset is augmented to 2112 volumes. \par

\subsection{Methodology}
% brief description of the flowchart
Our proposed framework is elucidated in Fig. \ref{fig:framework}. The system directly takes the whole US volume as an input. Our deep encoder-decoder architecture then densely labels each voxel in the volume, and generates intermediate probability maps for the foreground and background. Deep supervision is attached to boost training efficiency. Attention modules are injected in both upsample path and skip connections to form the hybrid filtering on volumetric features (Fig. \ref{fig:framework}(a)). Multiple segmentors then follow a cascaded fashion to refine the volume predictions level by level. Final output of the system is the segmentation of whole fetal head (Fig. \ref{fig:framework}(b)). \par

\subsection{3D Encoder-Decoder Architecture for Dense Labeling}
With the interleaved convolution, pooling and non-linearity layers to extract features and organize the semantic hierarchy, convolutional neural networks have become the dominant workhorse for medical image segmentation \cite{Shen_review}. Among all the architectures, FCN \cite{Long_fcn} with encoder-decoder design is a popular choice for dense pixel-wise end-to-end mapping. \par

Because 3D FCN has orders of magnitude of parameters than 2D ones, processing the whole US volume with 3D FCN is challenging under limited GPU memory. Previous researches often resort to slice \cite{Namburete2018fully,cerrolaza2018deep} and tri-plane based 2D FCN. However, they discard the spatial dependency and thus sacrifice accuracy \cite{dou2016automatic}. 3D patch-based FCN with an overlap-tiling stitching strategy is another attempt for volumetric segmentation \cite{yang2017hybrid}. Although it explores spatial cues in 3D patches, it is time-consuming and loses global context in the volume to regularize its segmentation. To directly process the whole US volume with 3D FCN, the architecture of 3D FCN should be carefully tailored. Also, all features should learn to be highly task-relevant. We will elaborate the details of our 3D FCN designs in this section. Improving the discriminative power of features with attention mechanism is introduced in the section follows. \par

As parameterized in Fig. \ref{fig:framework}(a), we customize a 3D U-net \cite{Ronneberger_unet} with long skip connections bridging encoder and decoder paths as our backbone. Concatenation operator is taken to merge feature volumes between encoder and decoder paths. Each convolutional layer (Conv) is followed by a batch normalization layer (BN) and a rectified linear unit (Relu). With the whole US volume as input, we focus on tuning the number of pooling layers, Conv layers and kernel sizes in our architecture to balance the input volume resolution and GPU memory constraints. All these factors affect the receptive field size and feature hierarchy of deep networks in perceiving global and local contexts. We finally opt for 2 successive Conv+BN+Relu layers as a block between each two max-pooling layers. There are totally 3 max-pooling layers in our encoder path. We select 3$\times$3$\times$3 kernel for shallow Conv layers, 5$\times$5$\times$5 kernel for deep Conv layers to further enlarge the receptive field. \par

% Deep supervision for 3D FCN
Suffering from the gradient vanishing problem along the long backpropagation path in deep networks \cite{Glorot_vanish}, deep layers are often over-tuned while the shallow layers are under-tuned during training. To maintain the training efficacy for all layers of our 3D deep network, we adopt the deep supervision mechanism to replenish the gradient flow with auxiliary losses and shorten the backpropagation path for shallow layers \cite{Lee-ds,dou2017_ds}. \par

As shown in Fig. \ref{fig:framework}(a), besides the main loss function at the end of network, deep supervision mechanism exposes shallow layers to the extra supervision of $\mathcal{M}$ auxiliary loss signals via $\mathcal{M}$ side branches. These branches share the same ground truth with the main loss function but have shortened backpropagation path length. These paths build a composite loss signal and therefore enhance the gradient flow in updating the parameters of shallow layers. The basic formulation of deep supervision is as follows:
\begin{equation}
\small
\label{eq:loss_ds_basic}
\begin{split}
\mathcal{L}(\mathcal{X},\mathcal{Y};W,w) = &\mathcal{L}(\mathcal{X},\mathcal{Y};W) + \\
&\sum_{m \in \mathcal{M}}{\mathcal{L}_m(\mathcal{X}, \mathcal{Y};W,w^m)} + \lambda(||W||^2),
\end{split}
\end{equation}
where $\mathcal{X}$, $\mathcal{Y}$ are training pairs, $W$ is the weight of main network, $w = (w^1,...,w^m)$, are the weights of side branches.

Several consecutive deconvolution operations are often used in the side branches to upsample the outputs as the same size as ground truth label \cite{dou2017_ds}. However, the deconvolution is very computation intensive and the number of it in these side branches is often times as that in the main network. Thus, this setting consumes a lot of GPU footprint. Recently, Lin \textit{et al.} \cite{lin2017refinenet} proposed to directly downscale the ground truth label to fit different branches and hence remove the heavy deconvolutions. As shown in Fig. \ref{fig:framework}(a), there is no deconvolution based upsampling in our side branches. We also downscale the ground truth label to 1/2 and 1/4 to fit the size of two branches, respectively. To increase non-linearity while keep computation cost, we use a Conv layer with 1$\times$1$\times$1 kernel to output the probability maps in each branch. Our modification preserves the effectiveness of deep supervision and makes more GPU memory available for the main network to explore. The final composite loss function for our deeply supervised network is accordingly modified as Eq. \ref{eq:loss_ds_resize}, 
\begin{equation}
\small
\label{eq:loss_ds_resize}
\begin{split}
\mathcal{L}(\mathcal{X},\mathcal{Y};W,w) = &\mathcal{L}(\mathcal{X},\mathcal{Y};W) +\\
&\sum_{m \in \mathcal{M}}{\mathcal{L}_m(\mathcal{X}, S_m(\mathcal{Y});W,w^m)} + \lambda(||W||^2)
\end{split}
\end{equation}
where cross entropy is the metric for main loss function $\mathcal{L}$ and auxiliary $\mathcal{L}_m$. $S_m(\cdot)$ indicates the downscale operation on ground truth label $\mathcal{Y}$. \par

\subsection{Hybrid Attention Scheme to Promote Features}
\label{ssc:attention}
Decoder path and skip connections assign the U-net variants advantages in distilling global context and preserving local details for fetal head segmentation. However, as the deconvolution successively upsamples the feature maps, not only the features representing the fetal head, but also the features of non-head region are learned at different scales. As these feature maps propagate along the decoder path, the final segmentation will be adversely affected. At the same time, feature maps at the shallow layers of encoder path not only contain the detailed boundary cues of fetal head, but also include abundant background noise. Skip connections convey these features to the decoder path to enhance the boundary details at the risks of bringing about false positives. Therefore, filtering the features to be task-relevant and focus on fetal head region becomes very important under the limited feature capacities and computation resource. \par

\begin{figure}[h]
	\centering
	\includegraphics[width=1.0\linewidth]{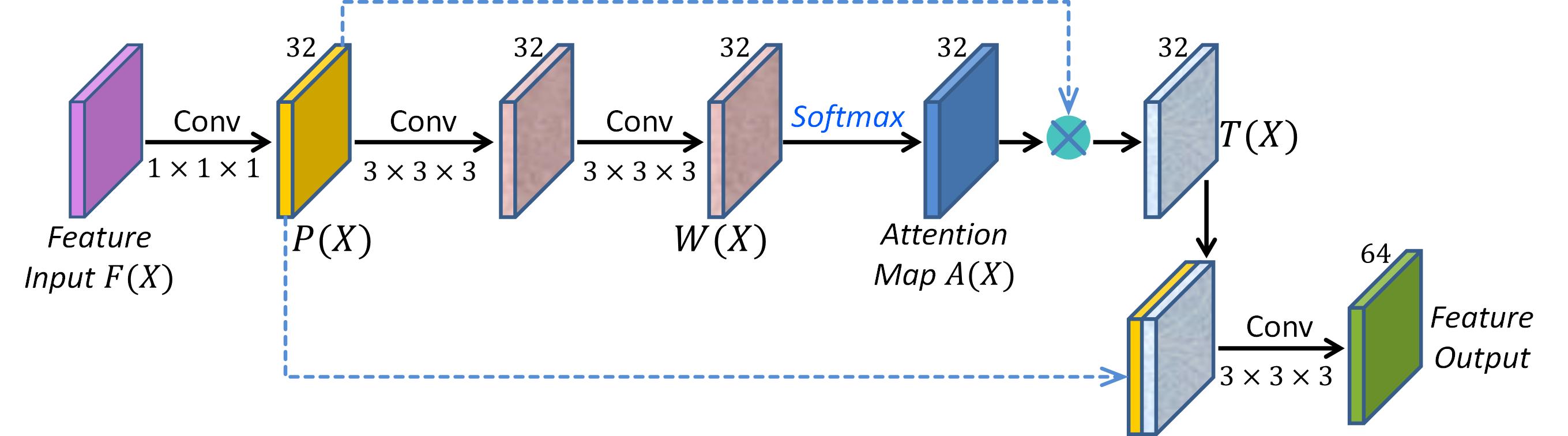}
	\caption{Schematic illustration of our proposed attention module.}
	\label{fig:attention_module}
\end{figure}

Attention mechanism becomes an attractive solution for the problem. It borrows the idea from human beings in applying the limited computation resources to the focus of view. Irrelevant features are suppressed under the mechanism to improve the learning efficiency and efficacy. Recently, attention mechanism becomes popular in image analysis \cite{wang2017residual} and natural language processing \cite{bahdanau2014neural}. In \cite{wang2018deep}, Wang \textit{et al.} proposed a framework with attention setting to improve multi-scale features for prostate segmentation in 2D US images. Schlemper \textit{et al.} further investigated the attention mechanism as a general module for both medical image classification and segmentation \cite{schlemper2019attention}. However, they only explored the attention module to filter the feature maps on skip connections and ignored the upsample path, and also the composite filtering effect. In this work, we propose a hybrid attention scheme (HAS) to progressively refine the features maps in both skip connection and upsample path at different scales. \par

As the building block of HAS, we firstly introduce the design of our attention module (AM) (Fig. \ref{fig:attention_module}). Given the feature maps $F(\mathcal{X})$ as input, we apply a Conv layer with 1$\times$1$\times$1 kernel to shrink the feature map channels to 32 as $P(\mathcal{X})$. This operation reduces the computation cost of our attention module and enhances the interactions among channels. We then attach two consecutive Conv layers with 3$\times$3$\times$3 kernel to produce the un-normalized attention weights $W(\mathcal{X})$:
\begin{equation}
\label{eq:attention_weights}
\small
W(\mathcal{X}) = \mathcal{H}(F(\mathcal{X});\theta)
\end{equation}
where $\theta$ represents the parameters learned by $\mathcal{H}$ which contains the mentioned 3 Conv layers. After that, we approach the attention map $A(\mathcal{X})$ by normalizing $W(\mathcal{X})$ across the channel dimension with a \textit{Softmax} function:
\begin{equation}
\label{eq:attention_softmax}
\small
a_{c,r,d}^{k}=\frac{exp(w_{c,r,d}^{k})}{\sum_{k}exp(w_{c,r,d}^{k})}
\end{equation}	
where $w_{c,r,d}^{k}$ denotes the value at spatial location $(c,r,d)$ and \textit{k}-th channel on $W(\mathcal{X})$, while $a_{c,r,d}^{k}\sim(0,1)$ denotes the normalized attention weight at spatial location $(c,r,d)$ and \textit{k}-th channel on $A(\mathcal{X})$. After getting the normalized attention map, we then multiply it with the $P(\mathcal{X})$ in an element-by-element manner to generate the refined feature map $T(\mathcal{X})$. With the attention mechanism, $\mathcal{H}(\theta)$ will learn to indicate and enhance the semantic features of fetal head boundary in $P(\mathcal{X})$ with higher $a_{c,r,d}^{k}$, while suppress the non-head regions with lower attention values (see our Results section). To fully make use of the $P(\mathcal{X})$ and $T(\mathcal{X})$, we concatenate them together and output the final 64-channel attention features after a Conv layer. \par

As shown in Fig. \ref{fig:framework}, to thoroughly exploit the feature filtering effect of attention mechanism, different from \cite{wang2018deep,schlemper2019attention}, our HAS implants the AM in both skip connections (denoted as \textit{SAM}) and upsample path (denoted as \textit{UAM}) at different scales in out network. HAS forms a composite and hierarchical feature selection in the segmentor. Specifically, we allocate a SAM in each skip connection and an UAM after each feature concatenation in the upsample path. UAM enforces the feature interaction among the concatenated feature maps and then selects the most discriminative features among them for further decoding. The auxiliary loss branches are attached behind the UAMs to ease the learning with filtered and discriminative features. \par

\subsection{Refinement with Auto-Context Scheme}
\label{ssec:refine}
Neighboring predictions are beneficial to support the decision on current location and address boundary ambiguity. Thanks to the whole US volume input and large receptive field size in our network design, our network can get arbitrary access to the context dependencies in long or short ranges. We hence combine our HAS based network with a classic iterative refinement framework, Auto-Context \cite{Tu_atoctxt}, to explore varying context and better recover the boundary of fetal head in ambiguous sites. \par

Auto-Context is designed to learn the context encoded in images and probability maps. It proves to be an elegant scheme for successive labeling refinement. As shown in Fig. \ref{fig:framework}(b), it stacks a series of models in a way that, the model at level $k$ can simultaneously revisit the appearance context in the intensity image and the shape context in the probability map which is generated by the model at level $k-1$. Eq.~\ref{equation2} illustrates the general iterative process of a typical Auto-Context scheme, 
\begin{gather}
\label{equation2}
\mathcal{Y}^{k} = \mathcal{G}^{k}(\mathcal{J}(\mathcal{X}, \mathcal{Y}^{k-1})),
\end{gather}
where $\mathcal{G}^{k}$ is the mapping function of the segmentor at level $k$, $\mathcal{X}$ and $\mathcal{Y}^{k-1}$ are the US volume and the probability map of fetal head from level $k-1$, respectively. $\mathcal{J}$ is a join operator to combine $\mathcal{X}$ and $\mathcal{Y}^{k}$, which is set as an element-wise summation in this work. Summation saves the computation cost and performs better than concatenation \cite{wu2017cascaded}. Limited by GPU memory, we train the network in level $k$ after finishing the training of level $k-1$. The probability map in level 0 is initialized with the constant value 0. By compromising between the time efficiency in training/testing and the performance gain, we use 2 context levels in total ($k=2$). The last context level outputs the final refined segmentation. \par

\section{Experimental Results}
\label{Results}
\subsection{Implementation and Evaluation Criteria}
We implemented our framework in \textit{Tensorflow}. Training and testing were run in a NVIDIA GeForce GTX TITAN X GPU (12GB). All the Conv layers were initialized from truncated normal distributions. As a trade-off between image quality and segmentation performance, we downscale the original US volume with a factor of 0.4 on each dimension for input. The final segmentation result was resampled back to the original full resolution for evaluation. We updated the weights of all layers with an Adam optimizer (batch size=1, initial learning rate=\textit{1e-5}, moment term is 0.5). The training epoch in each Auto-Context level was set to 30. For the testing time, our final method only needs about $2$ \textit{seconds} to segment an US volume. \par

% Evaluation metric
For segmentation evaluation, we target to assess the region, boundary and voxel-wise similarities with 5 criteria. They include the Dice similarity coefficient (DSC, $\%$), Conformity (Conf, $\%$), Jaccard (Jacc, $\%$), average distance of boundaries (Adb [mm]),  Hausdorff distance of boundaries (Hdb [mm]). DSC indicates the mutual overlap between segmentation and ground truth. Conformity ($Conf=(3DSC-2)/DSC$) provides wider range and can be more sensitive and rigorous than DSC, as suggested by \cite{Chang_Conform}. Adb is used to describe the average distance from segmentation surface to ground truth. Hdb is sensitive to boundary outliers and emphasizes the worst labeling cases \cite{Rueda_Challenge}. DSC, Jaccard, Adb and Hdb are defined as following,
\begin{equation}
DSC(E,G) = \frac{2|E\cap G|}{|E|+|G|}, Jacc(E,G) = \frac{|E \cap G|}{|E \cup G|},
\label{eq:dice, jaccard}
\end{equation}
\begin{equation}
\begin{split}
Adb(E,G) = &\frac{1}{2}(\frac{\sum_{v_{i}\in e_{G}}min_{v_{j}\in e_{E}}dist(v_{i},v_{j})}{|G|} +\\
&\frac{\sum_{v_{j}\in e_{E}}min_{v_{i}\in e_{G}}dist(v_{j},v_{i})}{|E|}),
\label{eq:adb}
\end{split}	
\end{equation}

\begin{equation}
\begin{split}
Hdb(E,G) &= max(H(e_{G}, e_{E}), H(e_{E}, e_{G})) \\
H(e_{G}, e_{E}) &= max_{v_{i}\in e_{G}}\left\{min_{v_{j}\in e_{E}}dist(v_{i},v_{j})\right\} \\
H(e_{E}, e_{G}) &= max_{v_{j}\in e_{E}}\left\{min_{v_{i}\in e_{G}}dist(v_{j},v_{i})\right\}.
\end{split}
\label{eq:hdb}
\end{equation}
where $E$ and $G$ are segmentation and ground truth. $|\cdot|$ calculates the volume of segmented object. $e_{E}$ and $e_{G}$ are the surfaces of segmentation and ground truth. $v_{i}$ is the vertex on the surface, $dist(v_{i},v_{j})$ is the Euclidean distance between vertex $v_{i}$ and $v_{j}$. \par

\subsection{Quantitative and Qualitative Analysis}
\label{quanti_quali}
Hence forth, we will denote our basic 3D segmentor network without deep supervision and HAS as \textit{USegNet}. As shown in Table \ref{table:quanti_diff_method}, we firstly compare the USegNet with several competitors, like the 3D deconvolution network (\textit{3D-DeconvNet}) \cite{noh2015learning}, 2D USegNet (\textit{2D-USegNet}) and 3D patch-based USegNet (\textit{p-USegNet}) to prove the effectiveness of our backbone architecture. The USegNet shares the same encoder-decoder layout with 3D-DeconvNet, 2D-USegNet and p-USegNet. Whereas, 3D-DeconvNet lacks the skip connections between encoder and decoder, 2D-USegNet takes slices with original resolution as input and outputs slices with the same sizes. p-USegNet digests 64$\times$64$\times$64 3D patches and generates the prediction of whole US volume with an overlap-tiling stitching strategy \cite{yang2017hybrid}. We keep proper settings for all the compared methods for fair comparisons. \par

\begin{table*} \caption{Quantitative comparison of different segmentation methods}
	\label{table:quanti_diff_method}
	\centering
	\tiny
	\resizebox{0.9\textwidth}{!}
	{
		\begin{tabular}{c|c|c|c|c|c}
			\toprule[2pt]
			\multirow{2}{*}{\bf{Method}} &\multicolumn{5}{c}{\bf{Metrics}} \\
			\cline{2-6}
			& DSC [\%] 		& Conf [\%]		& Jacc [\%]		& Adb [mm]		& Hdb [mm]		 	\\
			\hline
			p-USegNet 		&93.31		&85.53		&87.53		&0.8908		&7.839 	 		\\			
			2D-USegNet 		&94.31		&87.86		&89.27		&0.8023		&6.861		 	\\
			3D-DeconvNet 	&94.51		&88.31		&89.62		&0.6804		&6.334		 	\\			
			USegNet			&94.83		&89.06		&90.20		&0.6186		&5.400		 	\\
			USegNet-DS		&94.95		&89.33		&90.42		&0.6225		&5.409		 	\\
			\hline
			USegNet-DS-UAM		&95.57		&90.70		&91.54		&0.5242		&4.785		 	\\
			USegNet-DS-SAM		&95.63		&90.83		&91.64		&0.5362		&4.702		 	\\
			USegNet-DS-HAS		&95.85		&91.30		&92.05		&0.5070		&4.876		 	\\
			USegNet-DS-HAS-Ctx &\textbf{96.05}	&\textbf{91.74}		&\textbf{92.42}		&\textbf{0.4793}		&\textbf{4.609} 		\\
			\hline
			\toprule[2pt]
		\end{tabular}
	}
\end{table*}

Lacking of global contextual information for guidance, p-USegNet presents the worst results in terms of both shape and boundary similarities among the compared methods. With original whole slice as high resolution input and slice based large training dataset, 2D-USegNet gets better results than p-USegNet but still has high boundary distance errors suffering from the lack of spatial regularization. 3D-DeconvNet achieves 1.2 percent DSC improvements over p-USegNet and reduces the Adb error for about 15$\%$ over 2D-USegNet. This proves the importance of adopting 3D operators and taking whole volume as input to exploit global context to benefit whole fetal head segmentation. By establishing skip connections to revisit detailed boundary cues in multi-scale feature maps, our USegNet further refines the results over 3D-DeconvNet with 0.3 percent improvement in DSC. Finally, by adding the deep supervision (DS) to boost the training process, USegNet-DS gets another 0.1 percent improvement in DSC. \par

\begin{figure}[h]
	\centering
	\includegraphics[width=1.0\linewidth]{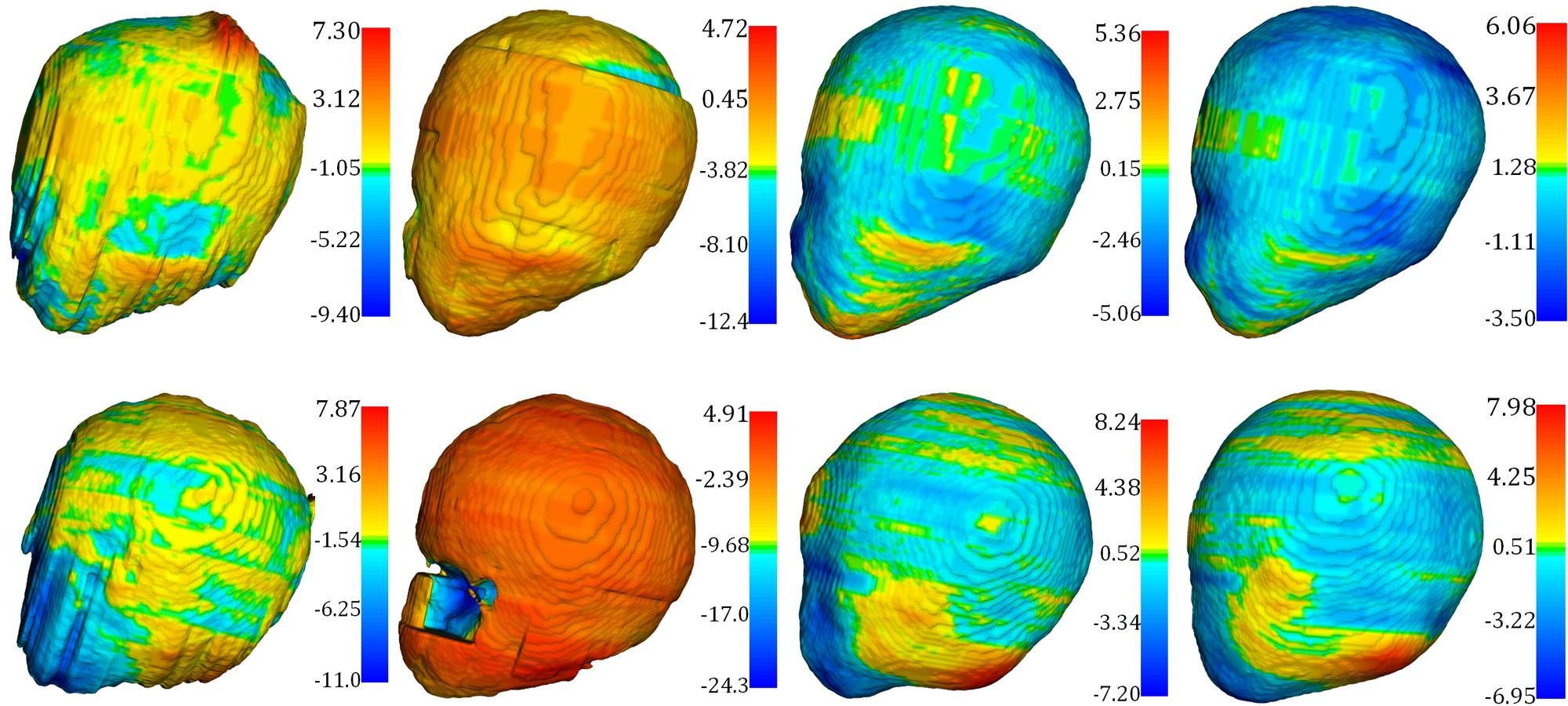}
	\caption{Two cases (first and second row) to show the comparison of Hausdorff distance [\textit{mm}] among different methods. From left to right: 2D-USegNet, p-USegNet, USegNet-DS-UAM and USegNet-DS-HAS-Ctx. The color bar is annotated with \textit{mean} in the center, \textit{min} and \textit{max} on the ends.}
	\label{fig:surf_dist}
\end{figure}

Based on USegNet-DS, we then move to conduct ablation study on our introduced modules, including the SAM, UAM, HAS and Auto-Context (Table \ref{table:quanti_diff_method}). Locating on the feature flow of main network, UAM (USegNet-DS-UAM) presents significant improvement over USegNet-DS, about 0.6 percent in DSC. Selectively enhancing the detailed boundary features of fetal head and discarding the noise in background as the upsampling progresses, UAM proves its importance in our architecture. Benefiting from the feature filtering effect on skip connections, SAM (USegNet-DS-SAM) brings  0.7 percent improvement in DSC and reduces the Adb error about 12$\%$ over USegNet-DS. This result reflects that, the feature maps from shallow layers of encoder path indeed contain redundant and irrelevant features. SAM suppresses these kind of features and hence improves the segmentation. Also, because UAM charges the main stream of feature flow in the network, UAM can bring higher improvements than SAM. \par

\begin{figure}[h]
	\centering
	\includegraphics[width=1.0\linewidth]{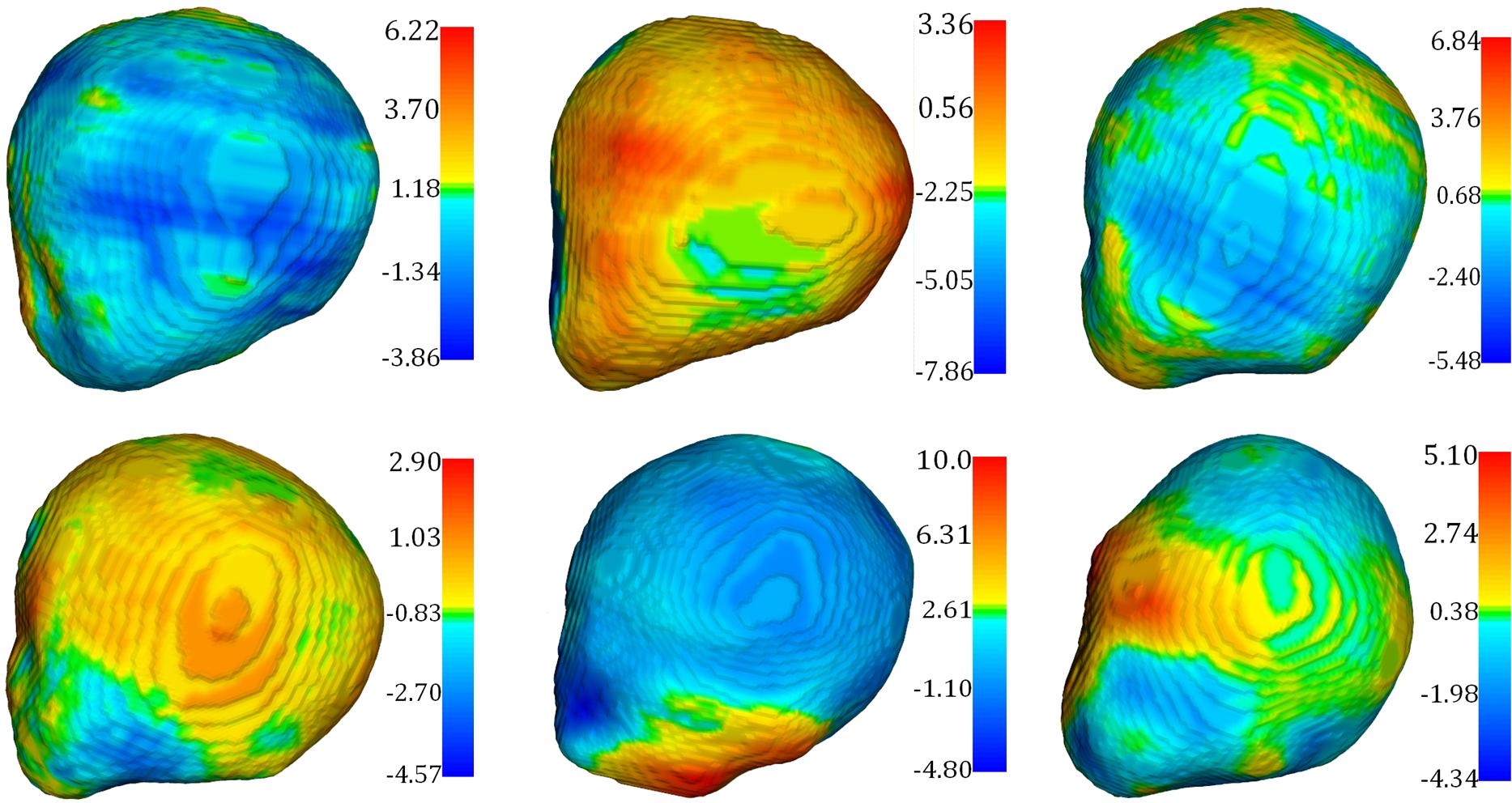}
	\caption{Segmentation result of six cases in the testing set. These cases have different shapes, sizes and gestational ages. Hausdorff distances [\textit{mm}] from the segmentation surface to ground truth are illustrated to provided a more detailed illustration. The color bar is annotated with \textit{mean} in the center, \textit{min} and \textit{max} on the ends.}
	\label{fig:attention_seg_comp}
\end{figure}

Combining the SAM and UAM to form the composite and hierarchical feature filtering, HAS (USegNet-DS-HAS) only adds little computation overhead but contributes more segmentation improvements than the SAM-only and UAM-only based models. It brings the highest refinement over USegNet-DS (about 0.9 percent in DSC, 19$\%$ in Adb). At this point, the whole fetal head segmentation performance of USegNet-DS-HAS is already very promising. The averaged absolute relative error in voxel number between our segmentation and ground truth gets as low as $3.40\%$. Considering the computation cost in training/testing and performance gain, our Auto-Context scheme only stacks two USegNet-DS-HAS with same configurations. We denote the stacked models as USegNet-DS-HAS-Ctx. As Table \ref{table:quanti_diff_method} shows, USegNet-DS-HAS-Ctx presents betterment on all metrics compared to USegNet-DS-HAS (about 0.2 percent in DSC, 0.4 percent in Jacc). Compared to the results reported by \cite{Namburete2018fully,cerrolaza2018deep,cerrolaza20183d} for fetal skull segmentation (as highest as 90$\%$ in DSC), our task is more challenging and our method  achieves better results. \par

% Visualization
With two cases as shown in Fig. \ref{fig:surf_dist}, we visualize the Hausdorff distance from different whole fetal head segmentation surfaces to the ground truth. Lacking of proper spatial context to guide and regularize the segmentation, the results of 2D-USegNet and p-USegNet are rough and visually implausible. The use of UAM (USegNet-DS-UAM) obviously reduces the surface distances of most boundary points. Benefiting from the hybrid feature filtering effect of HAS and Auto-Context, USegNet-DS-HAS-Ctx further narrows the surface distances. More visualization results of USegNet-DS-HAS-Ctx are shown in Fig. \ref{fig:attention_seg_comp}. It can be observed that, our proposed method conquers the poor image quality, scale and shape variations, occlusion and boundary ambiguities of whole fetal head in US volumes, and finally presents promising segmentation results. \par

\begin{figure}[h]
	\centering
	\includegraphics[width=0.95\linewidth]{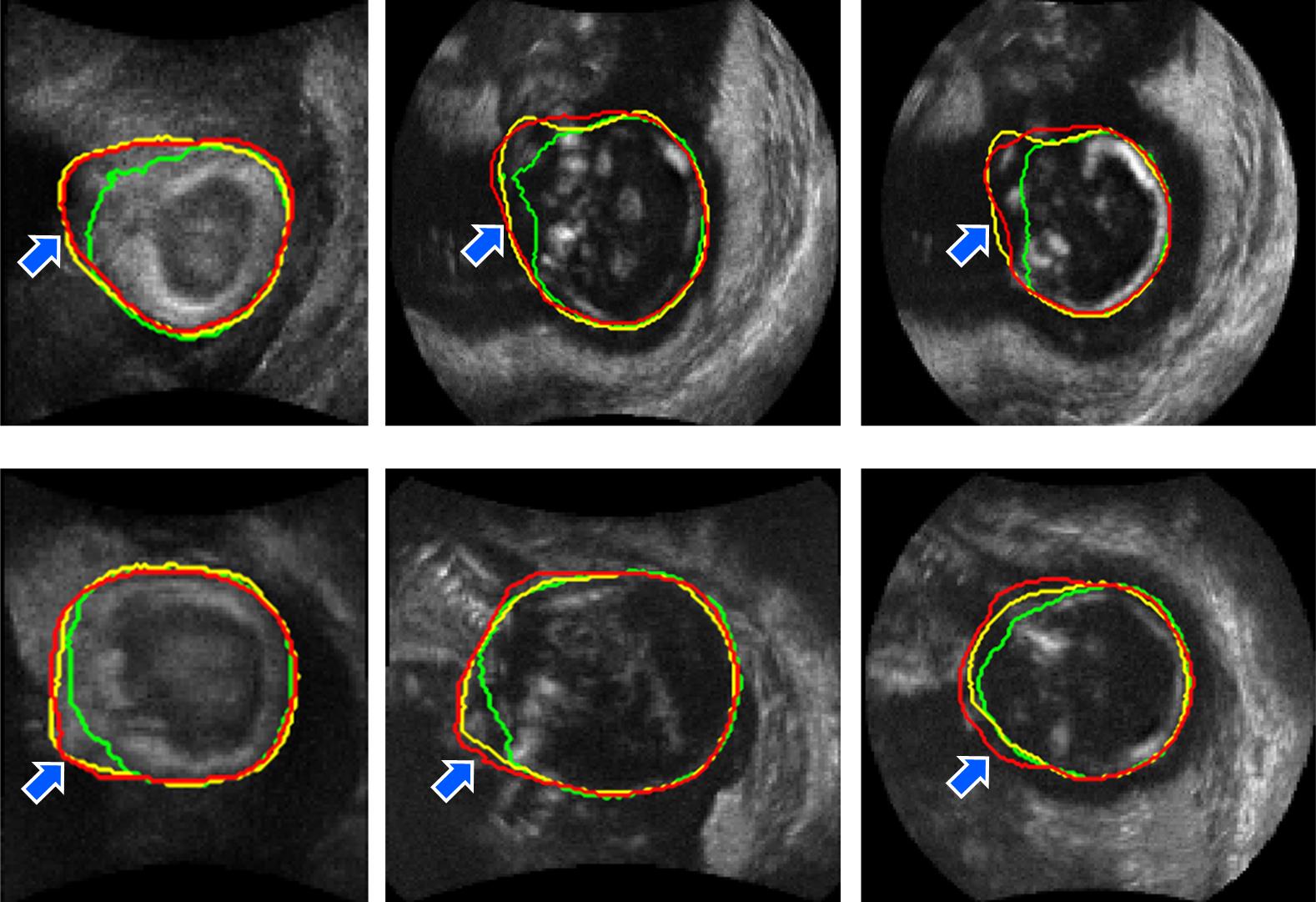}
	\caption{Two cases (first and second row) to explicitly compare the segmentation details of different methods. From left to right, sagittal slice of a whole fetal head in right, middle and left region. Red, green and yellow curves denote the contour from ground truth, USegNet-DS and our USegNet-DS-HAS.}
	\label{fig:seg_curve_comp}
\end{figure}

\begin{figure}[h]
	\centering
	\includegraphics[width=0.9\linewidth]{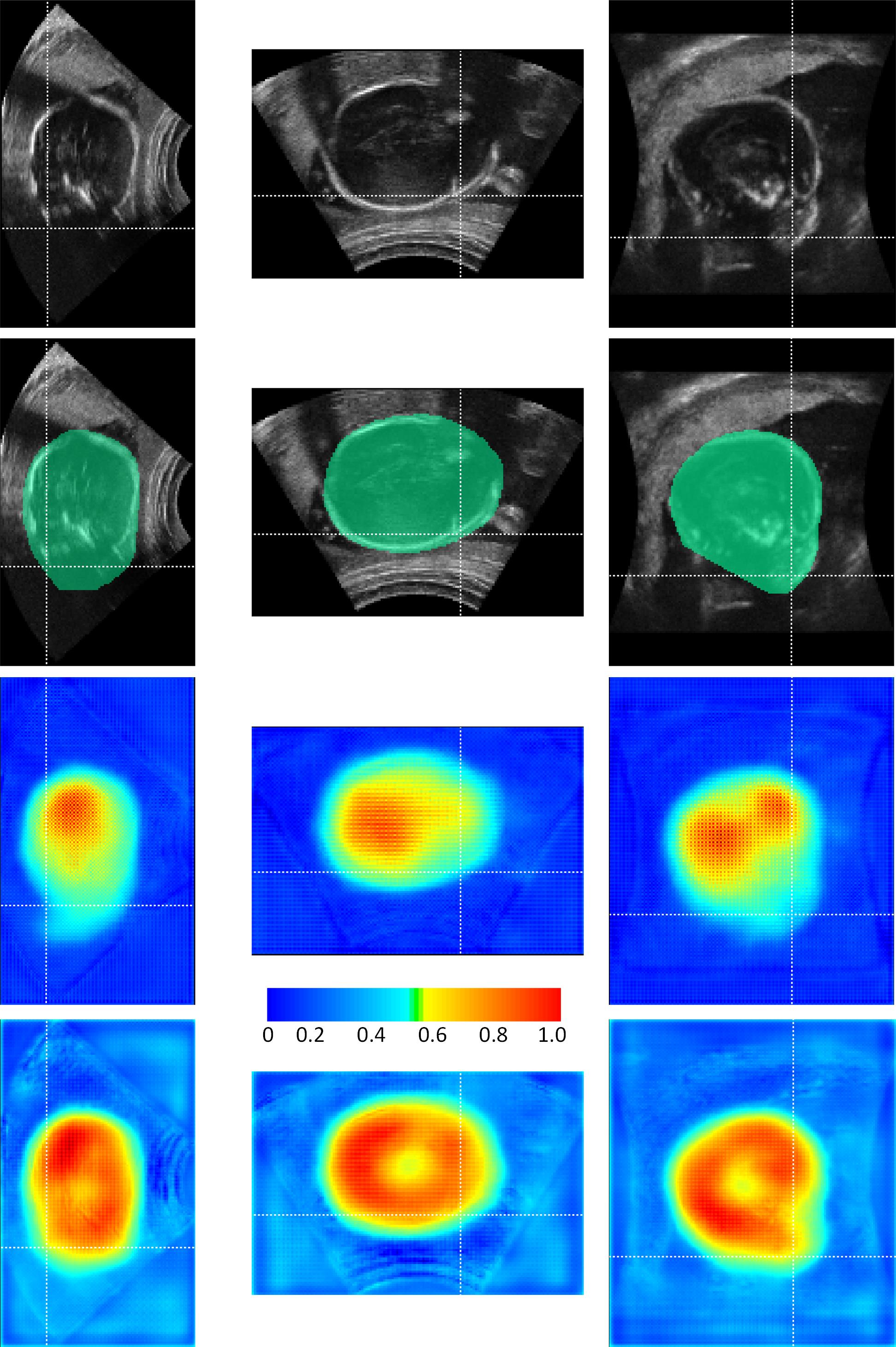}
	\caption{Advantages of HAS in enhancing the prediction maps. From top to bottom: coronal, traverse and sagittal slice of a whole fetal head; segmentation ground truth (green) overlaid on slices; probability map of whole fetal for three slices produced by USegNet-DS; probability map of whole fetal for three slices produced by USegNet-DS-HAS. Color map denotes the probability range. White crosshair is used to facilitate the point-to-point comparisons.}
	\label{fig:attention_map_comp}
\end{figure}

In Fig. \ref{fig:seg_curve_comp}, we illustrate an explicit comparison between USegNet-DS and our USegNet-DS-HAS to show the effectiveness of hybrid attention. As we can observe, both USegNet-DS and USegNet-DS-HAS can properly fit the ground truth around fetal skull regions. However, suffering from the irrelevant features in the background, USegNet-DS tends to under-segment the whole fetal head around the fetal facial and neck areas (blue arrows). The boundaries in these areas are hard to recognize due to the lack of hard bone structures. With Fig. \ref{fig:attention_map_comp}, we further show a case (USegNet-DS vs. USegNet-DS-HAS) to reveal the impact of our proposed hybrid attention. Through the point-to-point comparisons, we can see that, the probability maps produced by USegNet-DS are still fuzzy and low around fetal head boundaries. Whereas, the maps produced by USegNet-DS-HAS are compact and high around the whole fetal head, even in the severely occluded spots. This phenomenon demonstrates that our hybrid attention scheme not only suppresses the false positive prediction in non-head regions but also enlarges the gap between foreground object and background noise. \par

In Table \ref{table:SAM_UAM_Number}, we conduct the ablation study on the number of SAM and UAM. The base model for the experiments is USegNet-DS. Its results are listed for reference. As shown in Fig. \ref{fig:framework}, we define our experimental setting as follows. SAM-1 denotes the USegNet-DS only with the module SAM 1. SAM-12 denotes the USegNet-DS with both SAM 1 and SAM 2 modules in the skip connections. USegNet-DS-SAM is our model with all the 3 SAMs. UAM-1 denotes the USegNet-DS with only the UAM 1 module before the second deconvolution layer. UAM-12 denotes the USegNet-DS with both UAM 1 and UAM 2 modules. USegNet-DS-UAM is our model with all the 3 UAMs. For the experiments on UAM, if it is removed, then the auxiliary loss branch is attached to the concatenation layer before it to keep the training process to be fair. As we can see, using only one SAM and UAM, SAM-1 and UAM-1 can already bring obvious refinement on the segmentation (about 0.5 percent in DSC). Increasing the number of SAM and UAM also increases the performance. However, the increment is decreasing and the performance comes to a saturation in USegNet-DS-SAM and USegNet-DS-UAM. When compared with the computation cost of USegNet-DS, the SAM and UAM modules only add slight computation overhead. SAM-12 and UAM-12 consistently improve all the segmentation metrics over SAM-1 and UAM-1, except the Hdb. We interpret this phenomenon as that, SAM-2 and UAM-2 locate at the middle semantic levels and may miss some very detailed features that are only conveyed by SAM-3. They therefore present slight degradation in the strict metric Hdb which emphasizes worst boundary outliers. \par

\begin{table*} \caption{Comparison about different numbers of SAM and UAM}
	\label{table:SAM_UAM_Number}	
	\centering
	\scriptsize
	\resizebox{0.9\textwidth}{!}
	{
		\begin{tabular}{c|c|c|c|c|c|c}
			\toprule[2pt]
			\multirow{2}{*}{\bf{Network Layout}} &\multicolumn{6}{c}{\bf{Metrics}} \\
			\cline{2-7}
			& DSC [\%] 		& Conf [\%]		& Jacc [\%]		& Adb [mm]		& Hdb [mm]	&Time(s)		 \\
			\hline								
			USegNet-DS		&94.95		&89.33		&90.42		&0.6225		&5.409		&0.78 	\\
			\hline
			SAM-1	&95.44		&90.42		&91.30		&0.5556		&4.939		&0.76		  \\
			SAM-12	&95.60		&90.75		&91.59		&0.5380		&5.149		&0.78  \\
			USegNet-DS-SAM		&95.63		&90.83		&91.64		&0.5362		&4.702		&1.11  \\
			\hline
			UAM-1	&95.49		&90.51		&91.39		&0.5374		&4.832		&0.77  \\
			UAM-12	&95.54		&90.62		&91.48		&0.5350		&5.024		&0.81  \\
			USegNet-DS-UAM		&95.57		&90.70		&91.54		&0.5242		&4.785		&0.88  \\
			\hline
			\toprule[2pt]
		\end{tabular}
	}
\end{table*}

\begin{figure}[h]
	\centering
	\includegraphics[width=1.0\linewidth]{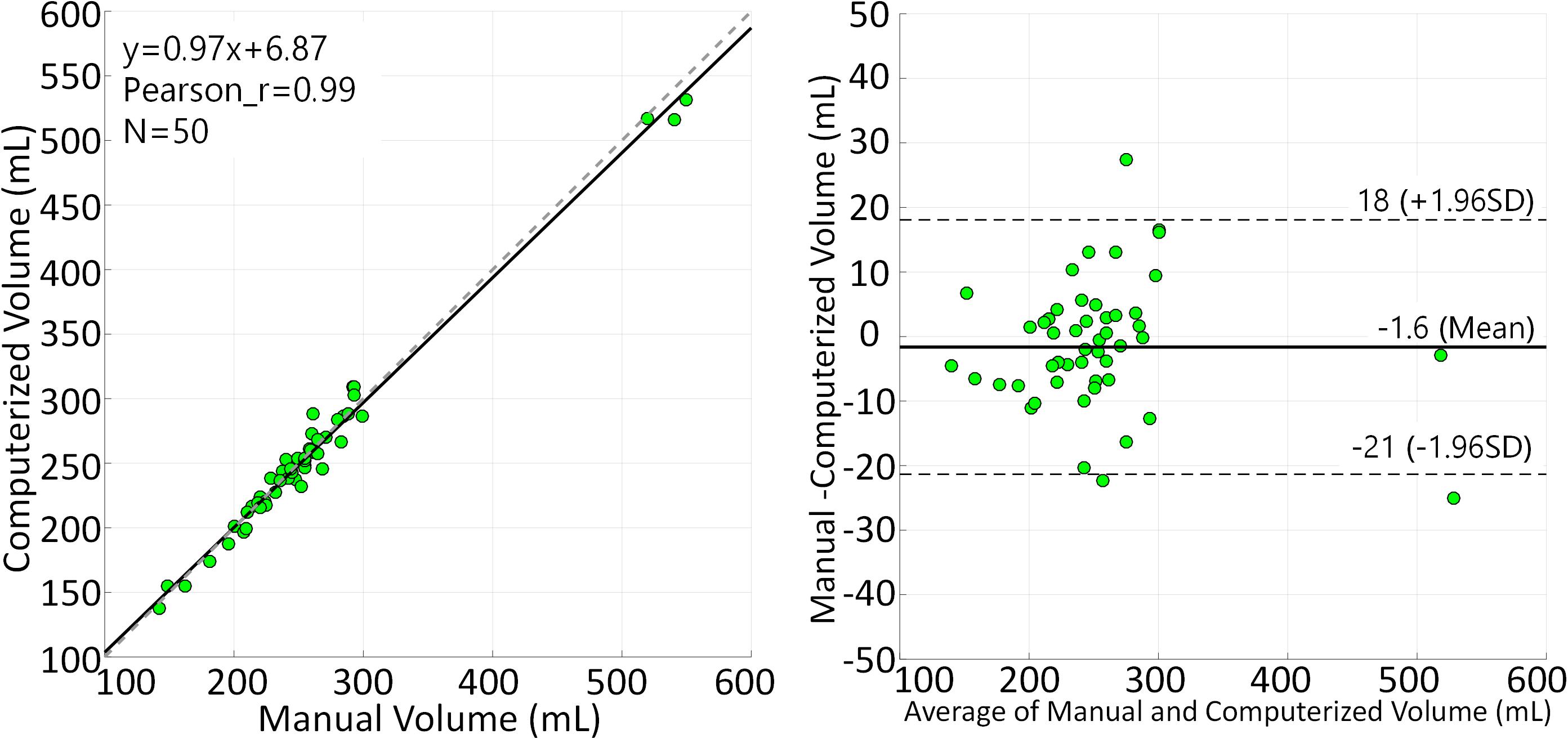}
	\caption{Correlation and Bland-Altman agreement on measuring the fetal head volume.}
	\label{fig:bland_altman}
\end{figure}

% Clinical test
After getting the fetal head segmentation, we can then obtain some useful biometrics, like the volume. We adopt the correlation coefficient and Bland-Altman agreement \cite{Rueda_Challenge} to comprehensively evaluate the discrepancy among the volume size derived from expert annotations and our USegNet-DS-HAS-Ctx segmentations. As shown in Fig. \ref{fig:bland_altman}, tested on the 50 varying volumes, our solution achieves high correlation (0.990) and agreement (-1.6$\pm$19.5 mL with 95\% of the measurements locate in the $\pm$1.96 standard deviation in Bland-Altman plot) in measuring the fetal head volume when compared to the expert. This high correlation and agreement indicate that our solution may serve as a promising alternative in assisting experts to analyze whole fetal head volumes. \par

\begin{figure*}[h]
	\centering
	\includegraphics[width=0.92\linewidth]{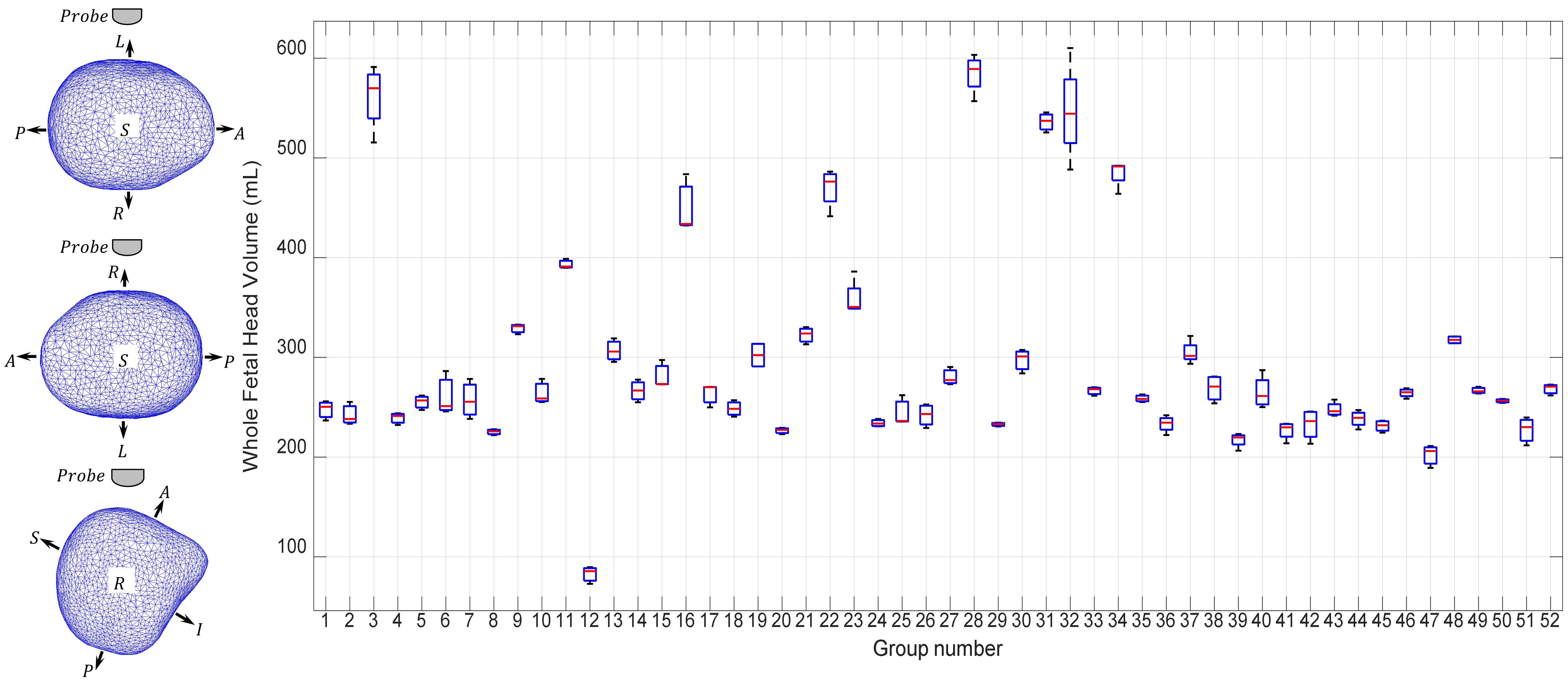}
	\caption{Fetal head volume (mL) measurement reproducibility (right) against three pre-defined scanning directions (left). Blue mesh represents the whole fetal head. Three volumes of a fetus in the same group.}
	\label{fig:vol_repro}
\end{figure*}

% Scanning variation
As shown in Fig. \ref{fig:challenges}, subject to the strong acoustic reflection on fetal skull, different fetal head orientations or scanning directions can arouse various shadows and occlusions. The appearance of whole fetal head in US volumes can hence drastically change. In this regard, keeping high reproducibility and being robust against scanning direction variation become crucial requirements before our methods can be applied in real clinical scenarios. Accordingly, we newly collected 156 volumes from 52 volunteers to validate the reproducibility of our solution (3 volumes per volunteer. Free fetal pose and varying GA from 21 to 31 weeks). The same US machine, S50 from SonoScape Medical Corp., Shenzhen, Guangdong Province, (China) was used for image acquisition. Each volunteer was scanned along three predefined directions, as shown in Fig. 9, where the Anterior(A), Posterior (P), Left (L), Right (R), Superior (S), Inferior (I) axes are sketched to denote fetal head orientation. For each volunteer, a volume is acquired along each direction. All the 3 volumes from the same volunteer are collected as a group. All the data are anonymized and the acquisition is approved by the local Institutional Review Board. All the volunteers have reviewed and signed the consent forms. Fig. \ref{fig:vol_repro} shows the box-plot of volume measurements generated by our USegNet-DS-HAS-Ctx for each group. We can observe that, our method suffers little from the pose or scanning variations, and attains remarkable reproducibilities (the mean of standard deviation of all groups is 11.524 mL, minimum is 1.960 mL, maximum is 49.869 mL) over all groups in measuring the whole fetal head volumes. \par

\section{Discussion}
\label{Discuss}
% Contribution, limitation and future work
Automated analyses of US volume have appealing potentials in promoting the prenatal examinations and bringing about changes to the traditional clinical work-flow. Accurately segmenting the whole fetal head in the volume may provide ever precise biometrics in describing the fetal growth. However, automated segmentation of whole fetal head in US volume is non-trivial due to the poor image quality, varying fetal poses and massive volume data. In this paper, we approach the task by proposing a fully-automated solution with high performances and good reproducibilities. \par

Whereas, there still exist several key points for future study. \textit{First}, since our method enables the automated extraction of fetal head volume, conducting the population study about the precise fetal head volume against GA becomes more tractable than ever. Previously, due to the lack of efficient tools in analyzing US volumes, there is no widely accepted reference chart of fetal head volume. This then limits the use of 3D US in supporting prenatal diagnoses. Only with the population study and the associated reference charts, volumetric measurement of fetal head can really benefit the fetal health monitoring. To achieve this goal, we need to collect more volume data and enhance our solution to cover a broad GA range. \textit{Second}, in the population study, ultrasound images will be acquired across different subjects, sites, devices, sonographers, GAs and etc. Unpredictable appearance shift in US images often happens during the acquisition due to different imaging conditions. Deep neural networks tend to suffer from this kind of appearance shift and be severely degraded \cite{yang2018generalizing}. Improving the generalization ability and robustness of deep neural networks to handle varying imaging conditions is critical for automated ultrasound image analysis, especially for the population study. Leveraging the shared shape prior \cite{yang2018generalizing} or fine-tuning the deep model for each acquisition site with few samples \cite{gibson2018inter} will be considered for our task. \textit{Third}, to accelerate the collection and annotation of large dataset for population study, we should greatly reduce the time and cost in manually annotating the volumes. Currently, the volume annotation is very expensive and time-consuming (more than 2 hours for one volume). Assisting the experts during annotation with machine learning powered algorithms, like the interactive segmentation \cite{wang2018interactive}, is highly demanded in our scenario. \textit{Finally}, based on our proposed automated segmentation, we should try to conduct longitudinal study to analyze the development pattern of fetal head volume. This kind of study may provide earlier and better indicators than 2D measurements for the prognosis of rare diseases, like the intrauterine growth restriction (IUGR) syndrome. \par

Considering the computation burden of the volumetric segmentation, we need to further reduce the computation cost of 3D deep networks to enable larger volume input with higher resolution for better segmentation results. We can consider the checkpointed backpropagation techniques \cite{chen2016training} to save more GPU memory for training with high resolution input. In \cite{yang2019fetusmap}, we explored the checkpointed backpropagation and provided clear evidences that higher resolution volumetric input can promote the localization of multiple fetal landmarks than low-resolution ones in 3D ultrasound. Real-time feature is not strongly required in current 3D ultrasound applications, however, network architectures like the pseudo-3D networks \cite{Qiu_pseudo3d} and Mobilenets \cite{howard2017mobilenets} to reduce the computation cost of convolution kernel should be seriously investigated in the near future. Also, as the fetal pose and scale vary greatly across subjects and timepoints, efficient detection strategies, like Faster R-CNN \cite{ren2015faster} in 2D or 3D form \cite{xu2019efficient}, to locate the fetal head in volume can greatly narrow the search space and hence reduces the computation burden in segmentation. Finally, the segmentation result should be fully explored to facilitate more advanced applications or be complementary to each other, like the landmark detection, standard plane localization and longitudinal analysis of fetal brain. \par

%%%%%%%%%%% Conclusions
\section{Conclusions}
\label{Conclusions}
In this work, we propose the first fully-automated solution for the precise segmentation of whole fetal head in US volumes. The task is pending and lacking satisfying solution before this work. We highlight our work with a hybrid attention scheme. It imposes a composite and hierarchical feature filtering effect on our 3D encoder-decoder backbone for better feature learning under the limited GPU resources and deep network layers. With experiments and demonstrations, our proposed modules are proved to be effective. Promising segmentation accuracy, remarkable correlations and agreements with experts, and high reproducibilities against scanning variations indicate that, our work may have potentials to assist sonographers in reviewing fetal growth from a new perspective. \par

\section{Acknowledgements}
%We would appreciate the great support from volunteers who participated in this study. 
This work was supported in part by the National Key R$\&$D Program of China (No. 2019YFC0118300), Shenzhen Peacock Plan (No. KQTD2016053112051497, KQJSCX20180328095606003), Medical Scientific Research Foundation of Guangdong Province, China (No. B2018031) and National Natural Science Foundation of China with Project No. U1813204. \par

\bibliographystyle{elsarticle-num}

\bibliography{cmpb_seg}
\end{document}